\documentclass[aps,prb,twocolumn,nopacs,floats,superscriptaddress]{revtex4-2}
\usepackage{lineno}       
\usepackage{hyperref}     
\usepackage{amsmath,amssymb} 
\usepackage{graphicx}     
\usepackage{float}        
\usepackage{soul} 
\usepackage{xcolor}
\sethlcolor{yellow} 
\begin{document}

\title{Bound states of $^3$He atoms at free $^4$He surfaces}
\date{March 2026}

\author{Massimo Boninsegni}
\affiliation{Department of Physics, University of Alberta, Edmonton, Alberta, Canada T6G 2H5}
\affiliation{Phenikaa Institute for Advanced Study, Phenikaa University, Nguyen Trac Street, Duong Noi Ward, Hanoi, Vietnam}

\begin{abstract}
Quantum Monte Carlo simulations confirm the existence of a bound state of a single $^3$He atom at a free superfluid $^4$He surface in three dimensions, localized within a $\sim$ 10 \AA\ thick low-density $^4$He surface layer,  with a binding energy of approximately 5 K  with respect to vacuum. As the temperature is raised above $\sim$ 1 K the $^3$He atom leaves the surface to dissolve into the superfluid. The situation is entirely different in two dimensions, as the $^3$He atom is excluded from the superfluid and no surface bound state exists. Results are also presented for a $^3$He atom binding to nanoscale size $^4$He clusters adsorbed on weak substrates, a physical system that may allow for the observation of some of the physics predicted for extended surfaces. It is shown that a $^3$He bound state localized at the perimeter of sufficiently small (a few tens of atoms) clusters exists not just on Cs, as previously reported, but on all alkali substrates.
\end{abstract}

\date{\today}
\maketitle
\section{Introduction}
A single $^3$He atom dissolves in superfluid $^4$He in the limit of temperature $T\to 0$, with a finite solubility limit of approximately 6.5\% at saturated vapor pressure. This has been experimentally known for a long time \cite{Graf1967,Edwards1969,Ebner1971}, and the energetics underlying this aspect of the phase diagram of the isotopic helium mixture is quantitatively understood \cite{Boninsegni1995,Ceperley1995}. Of special  interest is the physics of the mixture, in the limit of high $^3$He dilution, in the presence of a free superfluid $^4$He surface. It was first proposed by Andreev \cite{Andreev1966} that a bound state of a $^3$He atom localized at the $^4$He surface may exist; this was essentially an {\em ad hoc} hypothesis to account for some anomalous behavior of the surface tension of the isotopic mixture. Subsequent experimental work has confirmed the existence of such a bound state, with a binding energy (with respect to the vacuum) of the order of 5 K. More generally, $^3$He atoms confined to such surface states are expected to form a quasi two-dimensional Fermi gas with variable degeneracy, depending on the $^3$He concentration \cite{Edwards1978}. The interest in this subject has been recently rekindled by 
high-precision neutron scattering measurements imaging density profiles of $^3$He atoms at the surface of a thick film of a saturated $^3$He-$^4$He mixture \cite{Kirichek2024}.
\\ \indent
On very general grounds one might expect a $^3$He bound state to exist at a free $^4$He surface, as the interatomic potential is isotope-independent and the lighter $^3$He atom can reduce its kinetic energy by residing in the low-density transitional region between bulk $^4$He and vacuum.
Some theoretical understanding of the physical origin and character of the bound state came from variational \cite{Lekner1970,Saam1971,Krotscheck2008}  and density-functional \cite{Bashkin1995} studies. The latter has yielded a numerical estimate of the binding energy of the $^3$He atom to the free $^4$He surface of 5.2 K with respect to the vacuum, i.e., consistent with the experimental estimate. 
While these approaches offered valuable qualitative and semi-quantitative insight, they embody well-defined approximations and/or built-in assumptions — a variational ansatz in the former case, a semi-empirical energy functional in the latter — whose accuracy is difficult to assess in a controlled way. Moreover, they are both ground state methods, i.e., temperature effects which are potentially relevant for experimental observation are left out.
\\ \indent
It therefore seems worthwhile to revisit this long-standing problem using an alternative, robust methodology, namely finite-temperature Quantum Monte Carlo (QMC) simulations, whose only input is the microscopic many-body Hamiltonian, i.e., they are free of uncontrolled approximations. It is actually somewhat surprising that no QMC study has been published for this problem (i.e., for a single $^3$He atom in the vicinity of an infinite surface of superfluid $^4$He), despite the fact that virtually exact numerical estimates can be obtained for all the physically relevant quantities, including correlation functions, and also allow one to gain qualitative insight through direct,  visual inspection of many-particle configurations generated by the simulation. Aside from the ground sate study of a free superfluid $^4$He surface by Vall\'es and Schmidt \cite{Valles1988}, most of the QMC work on the physics of a single $^3$He atom in the presence of a free $^4$He surface has focused on thin $^4$He films adsorbed on different substrates \cite{Boninsegni2010,Boninsegni2022}, or at solid-liquid $^4$He interfaces \cite{Khairallah2005,Boninsegni2022b}. 
\\ \indent
This paper illustrates the results of a theoretical investigation, based on QMC simulations at finite temperature, of the occurrence of bound states of a $^3$He atom in the proximity of either extended or spatially confined $^4$He surfaces. The first case considered is that of an extended, free bulk superfluid $^4$He surface, for which the existence of a bound state is confirmed, localized into a $\sim$ 10 \AA-wide low-density surface layer with a binding energy of approximately 1.2 K with respect a $^3$He atom dissolved in bulk superfluid $^4$He, i.e., approximately 5 K with respect to vacuum, which is consistent with the estimate of Ref. \cite{Bashkin1995}. As the temperature is raised to above $\sim 1$ K, the $^3$He atom begins to leave the surface and move toward the interior of the bulk superfluid, as expected. Results are also presented for two $^3$He atoms floating on the $^4$He surface, showing that a single bound state exists, with a binding energy estimated at approximately 100 mK. To our knowledge, this is the first direct, parameter-free theoretical estimate of the binding energy of a ($^3$He)$_2$ dimer at a free $^4$He surface; it provides an order-of-magnitude assessment of the temperature at which a possible superfluid transition of a quasi-two-dimensional dilute $^3$He gas may undergo a superfluid transition of its own, giving rise to the intriguing, long sought scenario of two coexisting interacting superfluids.\\ \indent
The case is then considered of a strictly two-dimensional (2D) system, i.e., a semi-infinite $^4$He film with a one-dimensional free surface. The physical scenario in this case turns out to be very different from the three-dimensional one. First, the $^3$He atom does {\em not} dissolve in 2D superfluid $^4$He. Second, simulations carried out down to temperature $T=0.25$ K yield no evidence of a surface bound state for the $^3$He atom. In other words, reduction of dimensionality drastically alters the energetics and ensuing physical behavior of the system \cite{Stan1998}. 
 An interesting physical setting, which in some respects interpolates between the above two limits, is that of nanoscale size clusters of $^4$He atoms adsorbed on surfaces of different attractive strength, in the presence of a single $^3$He atom. The last part of this paper offers simulation results illustrating how the physics of the lone $^3$He impurity varies as the substrate becomes increasingly attractive, affecting the shape of the adsorbed $^4$He cluster. 

This paper is organized as follows: in section \ref{meth} the model and methodology utilized in this work are described; the results in presented in sec. \ref{res}, while in Sec. \ref{concl} the conclusions are outlined.

\section{Methodology} \label{meth}
All simulated systems comprise $N$ He atoms, one of them of mass $m_3$ being of the light isotope $^3$He and all others of $^4$He (of mass $m_4$); they are all enclosed inside a $L\times L\times L_z$ parallelepipedal cell ($L\times L_z$ in the case of a 2D system) with $L_z > L$ (precise values are given below). Periodic boundary conditions are utilized in all directions. 
The two opposing faces of the cell along the elongated direction (i.e., $z=0$ and $z=L_z$) represent the ends of semi-infinite, homogeneous slabs of matter, either superfluid $^4$He at equilibrium density, or a substrate on which helium can be adsorbed. Unless otherwise specified, the interaction of the He atoms with these end surfaces is described in three dimensions via suitably parametrized `3-9'' potentials
\begin{equation}\label{39}
    V(z) =\frac{D}{2}\ \ \biggl [ \biggl ( \frac {a}{z}\biggr )^9-3\ \biggl (\frac{a}{z}\biggr )^3\biggr ]\ \ ,
\end{equation}
In two dimensions, the above expression is replaced by the equivalent ``4-10'' one
\begin{equation}\label{410}
    V(z) ={D}\ \ \biggl [\frac{2}{3} \biggl ( \frac {a}{z}\biggr )^{10}-\frac{5}{3}\ \biggl (\frac{a}{z}\biggr )^4\biggr ]\ \ .
\end{equation}
$D$ is the depth of the attractive well of the interaction and $a$ the distance of closest approach, below which the atom experiences a strong (hard core) repulsion.
The values of $D$ and $a$ are chosen  to represents the specific medium considered. For example, for bulk superfluid $^4$He in three dimensions at the $T=0$ equilibrium density $\rho_{3d}=0.021837$ \AA$^{-3}$ it is $D=8.226$ K, $a=2.2$ \AA\ (``3-9'' potential), while in two dimensions ($\rho_{2d}=0.0432$ \AA$^{-2}$) it is $D=2.7$ K and $a=2.38$ \AA\ (``4-10'' potential). Additional details are provided in section \ref{res} outlining the results for the different physical settings considered.
The interaction between two He atoms is described by the accepted Aziz pair potential \cite{aziz1979}.
\\ \indent
Thermodynamic properties of the system have been computed by QMC simulations at finite temperature ($T$), based on the canonical variant \cite{Mezzacapo2006,Mezzacapo2006b} of the continuous-space Worm Algorithm \cite{Boninsegni2006,Boninsegni2006b}. Details of the simulation are  standard; a short-time propagator was used which is accurate to fourth order in the time step $\tau$ \cite{Jiang2001,Boninsegni2005}. Numerical extrapolation of the estimates to the $\tau\to 0$ limit show convergence of the thermal averages for a value of $\tau=3.125\times 10^{-3}$ K$^{-1}$ for all quantities of interest here. These include the energy per particle, $^4$He density profiles and $^3$He probability density of position.

\section{Results}\label{res}
\subsection{Free superfluid $^4$He surface in three dimensions}
In the simulation described in this subsection, $N=1025$, $L=28.623$ \AA\ and $L_z=306$ \AA. The two walls at the end of the long side of the cell describe bulk superfluid $^4$He. The $^4$He atoms are initially arranged on a simple cubic lattice built on top of the $z=0$ wall, with a density $\rho_{3d}$, extending up to $z\approx 60$ \AA. The single $^3$He atom is initially placed slightly above the top $^4$He atom layer.\\ \indent
\begin{figure}[h]
\centering
     \includegraphics[width=1.0\linewidth]{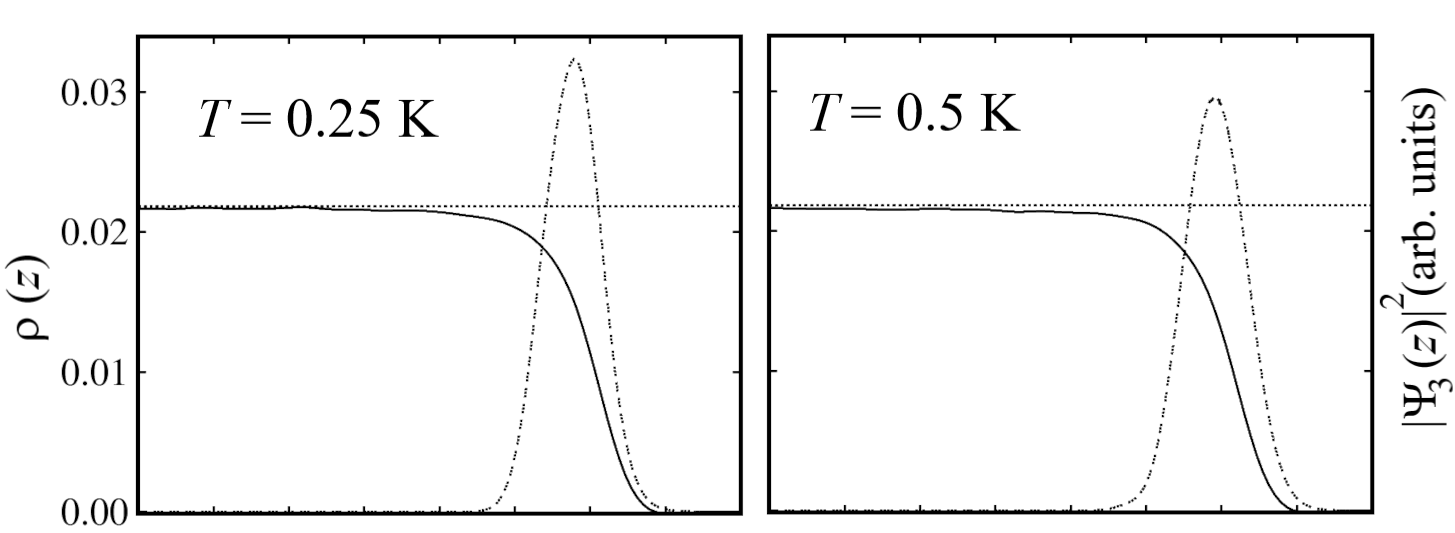}
          \includegraphics[width=1.0\linewidth]{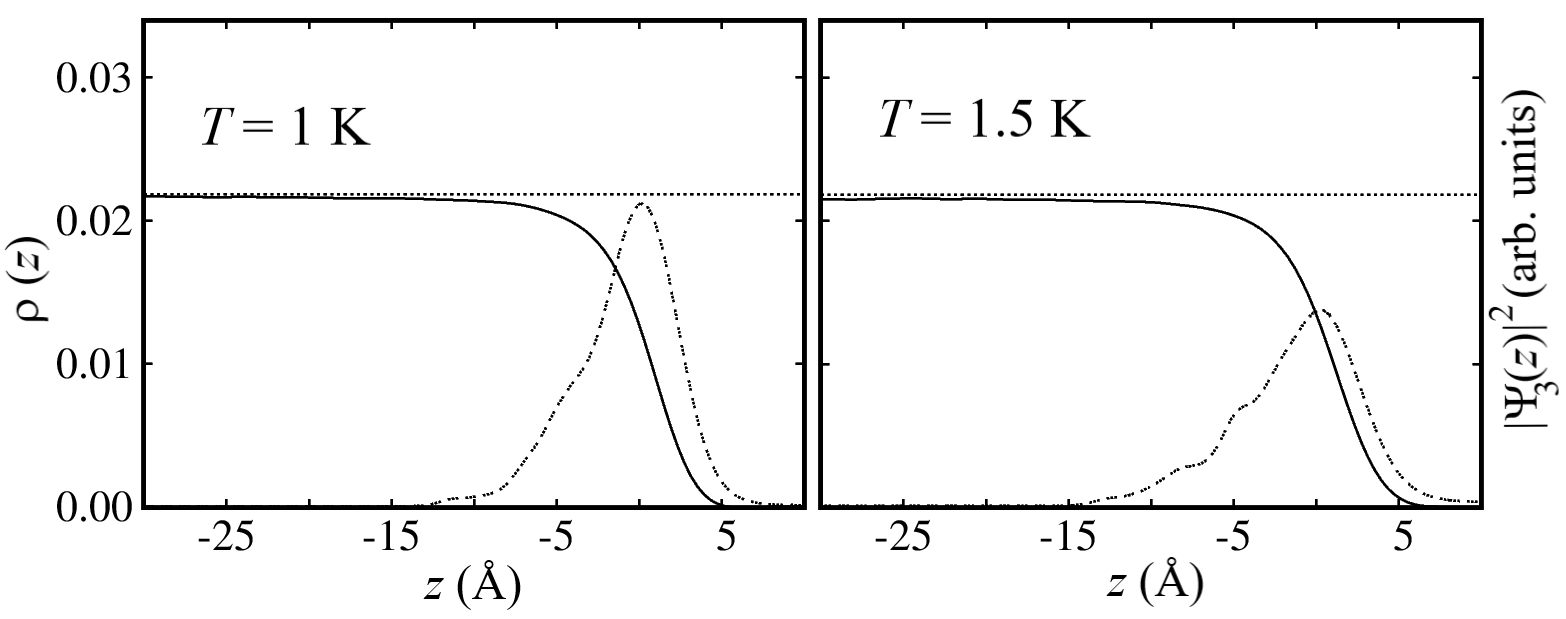}
\caption{$^4$He density profile $\rho(z)$ (in \AA$^{-3}$, solid lines) plotted along the direction perpendicular to the free surface at four different temperatures. The location of the origin is arbitrary, the same for all temperatures. Also shown (dashed lines) is the probability density of position for the $^3$He atom (arbitrary units). Horizontal dotted line shows the $T=0$ equilibrium density $\rho_{3d}$ of superfluid $^4$He. \label{profiles}}
\end{figure} 
The relatively large elongation of the cell in this case is aimed at making periodic boundary conditions along the $z$ direction irrelevant, allowing for the unambiguous observation of a free surface of $^4$He, formed by the actual $^4$He atoms; the potential associated to the wall at $z=0$ has the sole purpose of accounting for the presence of bulk superfluid $^4$He in the $z < 0$ region, in an effective way. The trade-off with adopting this geometry is the relatively small cross-sectional area of the system (i.e., in the $xy$ plane), which sets an upper cutoff for the wavelength of surface ripplons, possibly affecting some of the physics of interest here.
Still, the cross-sectional area in this study is slightly greater than in the original QMC study of a free superfluid $^4$He surface \cite{Valles1988}, which provided an altogether reasonably satisfactory quantitative account of its basic physics. The system size utilized is sufficient to ensure no significant evaporation of $^4$He atoms.
\\ \indent
Fig. \ref{profiles} shows the computed $^4$He density profiles $\rho(z)$ at four different temperatures, the lowest being $T=0.25$ K, the highest $T=1.5$ K. A clear, reasonably sharp surface is observed, with $\rho(z)$ going from essentially zero to a value close to $\rho_{3d}$ within a $\sim$ 10-\AA\ distance. There is a small, although noticeable dependence of the result on the temperature, with the surface spreading out further (by few \AA) at the highest temperature, with the concurrent, progressive lowering of the equilibrium density.  Altogether, the result shown in the figure gives an idea of the finite size effects affecting the original ground state QMC calculation of Ref. \cite{Valles1988} (carried out on a system comprising an order of magnitude fewer atoms), yielding a surface width of about 2 \AA. Better quantitative agreement is observed with the Density Functional study of Dalfovo {\em et al.} \cite{Dalfovo1995}.
\\ \indent
Also shown in Fig. \ref{profiles} is the probability density of position of the $^3$He atom $|\Psi_3(z)|^2$ (dashed lines), in arbitrary units. In  the $T\to 0$ limit the $^3$He atom remains confined to the low-density $^4$He surface layer, in a way that seems very similar to what has been predicted in the case of relatively thick (more than 10 layers) $^4$He films adsorbed on weakly attractive (alkali) substrates; in fact, even the free $^4$He surface determined in this calculation seems virtually indistinguishable from that obtained for a $^4$He film of coverage 0.06 \AA$^{-2}$ adsorbed on Cs or Li \cite{Boninsegni2022}.
\\ \indent
The binding energy of the $^3$He atom to the surface can be estimated as outlined in Ref. \cite{Ceperley1995}.
The chemical potential $\mu_3$ of a single $^3$He atom dissolved in superfluid $^4$He at $T=0$ is given by
\begin{equation}\label{ti}
\mu_3=\mu_4+\int_{\lambda_4}^{\lambda_3} d\lambda\ \frac{K(\lambda)}{\lambda}
\end{equation}
where $\mu_4$ is the $T=0$ chemical potential of superfluid $^4$He atoms at the equilibrium density $\rho_{3d}$, $\lambda\equiv \hbar^2/2m$, $m$ being the mass of a single atom that is ``tagged'' (i.e., regarded as distinguishable from the others), $K(\lambda)$ is the kinetic energy of the distinguishable atom as its mass is varied continuously, and the integration limits are between the values of $\lambda$ which corresponds to a $^4$He ($\lambda_4$) and a $^3$He ($\lambda_3$) atom. 
An estimate for $K(\lambda_3)$ was obtained in this work
by performing a simulation for bulk $^4$He with a single
$^3$He atom down to temperature $T = 0.25$ K, yielding $K(\lambda_3) = 17.08(3)$ K; using the most current estimates for $K(\lambda_4), \mu_4$, namely 
$K(\lambda_4) = 14.126(2)$ K, 
$\mu_4=-7.233(2)$ K \cite{boninsegni2026}, we obtain $\mu_3=-2.817(5)$ K.
\\ \indent
Thermodynamic integration over the impurity mass (Eq. \ref{ti}) was carried out piecewise, by performing a few simulations in which the value of $\lambda$ for the foreign atom was varied in steps of 0.5 K\AA$^2$, also using the concavity of the free energy with respect to $\lambda$ to bound the integral. The computed kinetic energy of the $^3$He atom at the surface is 8.3(1) K; we find the ratio $K(\lambda)/\lambda$ to decrease monotonically, fairly rapidly as $\lambda$ is increased from $\lambda_4$ to $\lambda_X$, with $X\approx 6.3$ K\AA, and then more slowly from $\lambda_X$ to $\lambda_3$, yielding a $^3$He surface binding energy $\Delta = 4.8(2)$ K. An independent estimate from the decay of the impurity Green function $G(k = 0, \tau)$ (see below), fitted to a single exponential yields $\Delta =6.0(6)$ K. Both estimates are consistent with the theoretical result from Density Functional Theory \cite{Bashkin1995}, as well as with the experimental determination \cite{Edwards1978}.  \\ \indent
This value of $\Delta$ is consistent with the behavior observed here as a function of temperature, shown in the four panels of Fig. \ref{profiles}. Specifically, as the temperature is raised the probability density $|\Psi_3(z)|^2$ becomes less peaked and more spread out, and the atom progressively moves toward the interior of the bulk superfluid for $T \gtrsim 1$ K, in what would normally be regarded as ``evaporation'', except that in this case it takes place with the atom not moving into vacuum but in superfluid $^4$He.

\subsection{Free superfluid $^4$He surface in two dimensions}
It is interesting to explore the 2D case of the same problem, i.e., the existence of a bound state of a $^3$He atom at the (linear) surface of a 2D $^4$He film. This physical limit can presumably be approached by considering a quasi-2D superfluid $^4$He film adsorbed on a substrate sufficiently strong to stabilize it but not too strong to cause atom localization by corrugation \cite{Boninsegni1999}. There have been some theoretical studies that have been based on the assumption that a homogeneous isotopic helium mixture is thermodynamically stable in two dimensions \cite{Krotscheck2001}.
\begin{figure}[h]
\centering
     \includegraphics[width=1.0\linewidth]{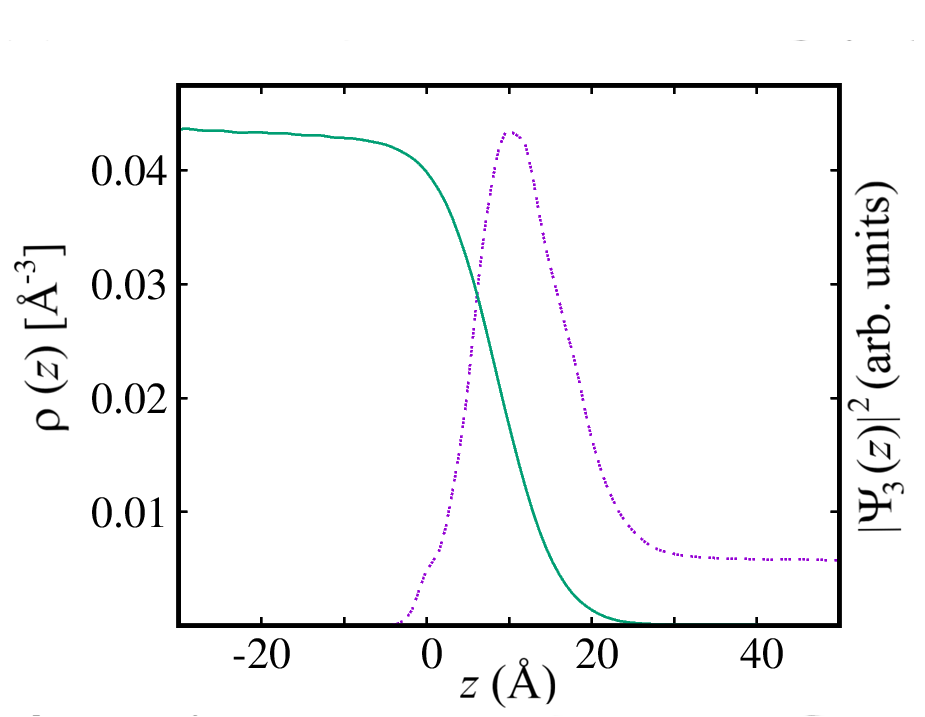}
\caption{$^4$He density profile $\rho(z)$ (in \AA$^{-2}$, solid line) plotted along the direction perpendicular to the free surface. The temperature is $T=0.25$ K. The location of the origin is arbitrary. Also shown (dashed line) is the probability density of position for the $^3$He atom (arbitrary units). \label{2prof}}
\end{figure} 
\\ \indent
As shown below, however, lowering of the dimensionality drastically alters the physics of the system. The first indication comes from the fact that, unlike in three dimensions, a single $^3$He atom will {\em not} dissolve in 2D superfluid $^4$He, a fact which does not appear to have been noticed or investigated so far. A calculation carried out in this work for 2D $^4$He at the 2D equilibrium density $\rho{2d}$ with a single $^3$He atom, yields a ground state energy per $^4$He atom of $-0.851(9)$ K, in agreement with previous work \cite{Ceperley1989},  with a kinetic energy per $^4$He atom equal to $3.928(9)$ K (also in agreement with the same reference) and for the $^3$He atom of 4.11(4) K. Straightforward application of Eq. \ref{ti} yields a {\em positive} value for $\mu_3$ in 2D, i.e., it is energetically unfavorable for a $^3$He atom be dissolved in superfluid $^4$He in two dimensions. Furthermore, the numerical solution of the one-dimensional Schr\"odinger equation for a single $^3$He atom moving in the presence of the external potential of the form (\ref{410}), with the parameters for 2D $^4$He specified above, fails to yield a bound state, i.e., one ought not expect even a weakly bound state with the $^3$He atom loosely floating atop the $^4$He film.
\\ \indent
In order to confirm this prediction, a QMC simulation was carried out based on the same setup used in three dimensions, this time with $N=576$, $L=115.47$ \AA, $L_z=500$ \AA. The simulation was carried out at a temperature $T=0.25$ K, which is sufficiently low to prevent significant evaporation; this is known to be a problem with $^4$He in reduced dimensions, given the weakness of the interatomic potential and the strong tendency of atoms to escape into the vacuum (unlike other types of 2D systems, e.g., parahydrogen, which crystallize and possess no metastable fluid phase at low $T$ \cite{Boninsegni2004}).
\\ \indent
Fig. \ref{2prof} shows the corresponding $^4$He density profile and $^3$He probability density of position, at temperature $T=0.25$ K. The obvious qualitative difference with respect to the result for the three-dimensional case, is that there is no evidence of binding of the $^3$He atom to the free $^4$He linear surface. Although $|\Psi_3(z)|^2$ has a significant enhancement at the surface, and in that region displays some of the same features as its 3D counterpart, its flat tail at long distance is conclusive evidence that no bound state exists. This conclusion seems robust, even though the calculation is at finite temperature; for, if a bound state existed with binding energy $<$ 0.25 K, some (possibly slow) decay of $|\Psi_3(z)|^2$ ought to be seen, if the binding energy were close to 0.25 K, otherwise nearly complete evaporation should occur, with no similar enhancement near the surface.
\subsection{$^4$He nanodroplets on alkali substrates}
Physically realizable systems exist that might allow one to observe experimentally the dramatic effect of dimensional reduction illustrated above. While the purely 2D scenario illustrated above is clearly of academic interest only, it can nonetheless be experimentally approached, e.g., by depositing helium nanodroplets on sufficiently weak substrates, such as those of alkali metals \cite{Cheng1993}. 
\begin{figure}[h]
\centering
     \includegraphics[width=1.0\linewidth]{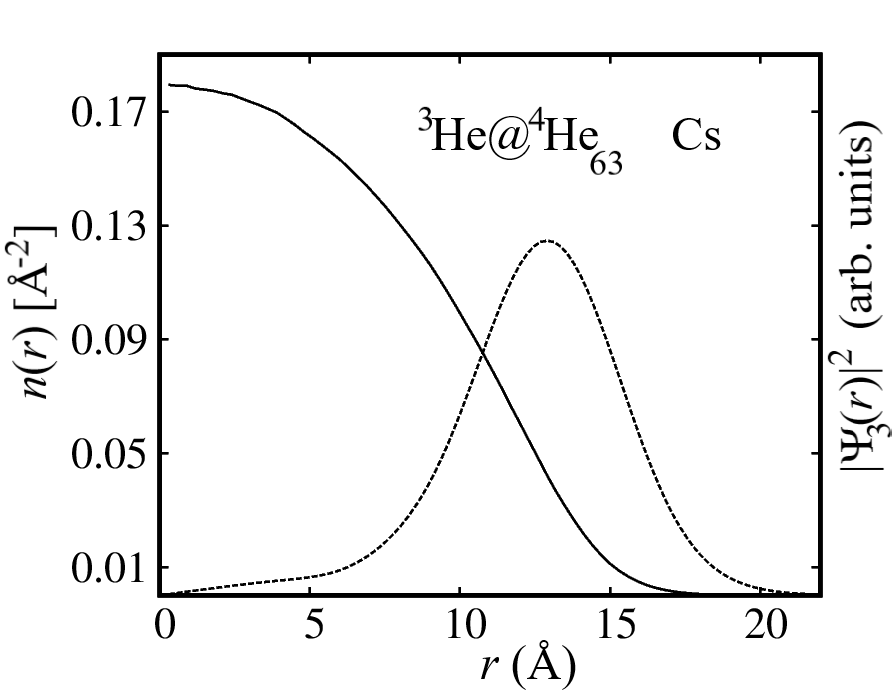}
\caption{Radial $^4$He density profile $n(r)$ (in \AA$^{-2}$, solid lines) integrated along the direction perpendicular to the substrate, for a cluster of $N=64$ helium atoms (one of $^3$He, the others $^4$He) adsorbed on a Cs substrate. Profile is plotted with respect to an axis through the center of mass of the cluster. Also shown (dashed line) is the probability density of position for the $^3$He atom (arbitrary units). \label{dropCs}}
\end{figure} 
\begin{figure}[h]
\centering
          \includegraphics[width=1.0\linewidth]{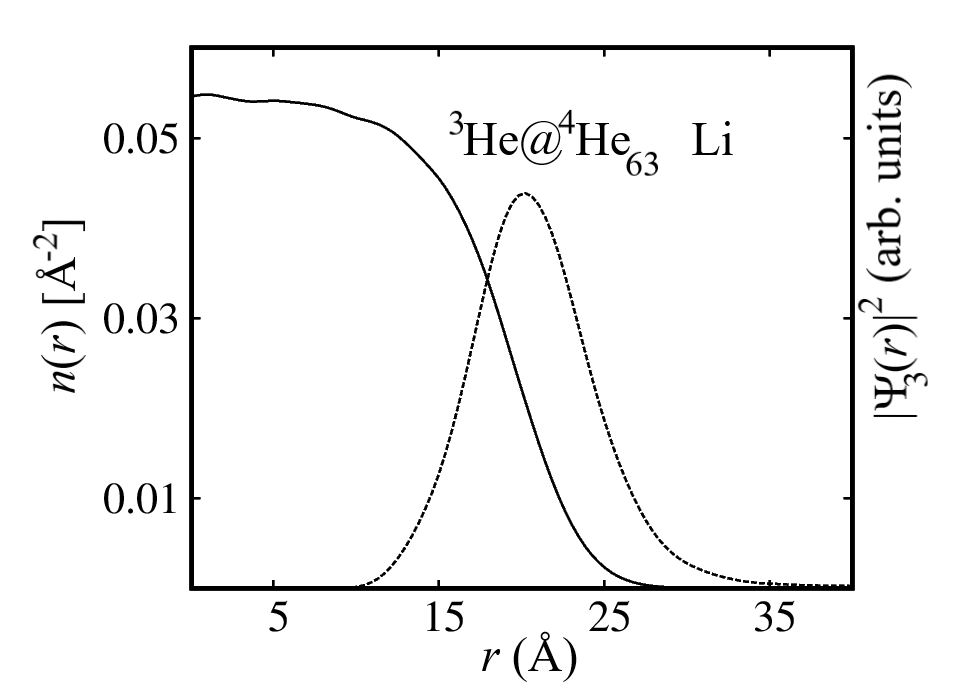}
\caption{Same as Fig. \ref{dropCs} but for a Li substrate. \label{dropLi}}
\end{figure} 
\begin{figure}[h]
\centering
          \includegraphics[width=1.0\linewidth]{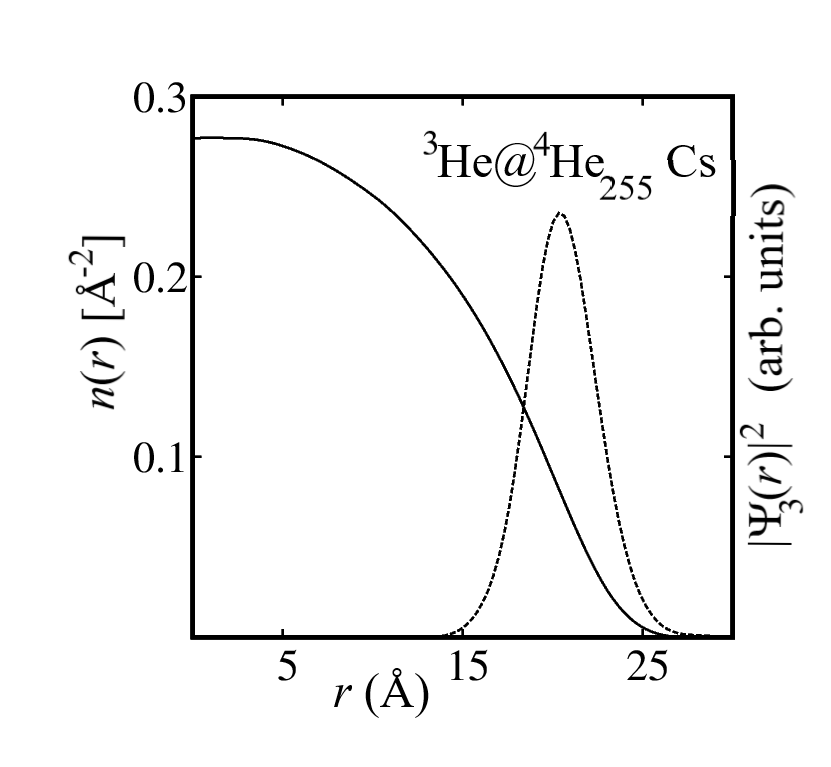}
\caption{Same as Fig. \ref{dropCs} but for a cluster of $N=256$ atoms. \label{dropCs2}}
\end{figure} 
In general, the corrugation of the substrate significantly affects the physics of a small cluster of helium atoms adsorbed on it \cite{Kolevski2025}. On the other hand, the effect of corrugation is nearly negligible in the case of a weak substrate, which can be regarded as nearly ``flat''. As a result, a small cluster of helium atoms adsorbed on it is floppy, liquid-like, and takes on a different shape depending on the substrate strength. For the case of Cs, which is known not to be wetted by helium, the shape of the nanodroplet is essentially that of a portion of a sphere, whose physical properties are to some extent affected by curvature \cite{Hernandez2003}. On the other hand, the droplet becomes flatter, resembling a pancake, on more attractive substrates such a Li, which is the weakest on which a superfluid $^4$He monolayer forms \cite{Boninsegni1999,VanCleve2007}. In this sense, one can meaningfully expect the physical properties of the system to evolve from  3D to (quasi) 2D.
\\ \indent
Simulations at temperature $T=0.125$ K were carried out for clusters on $N=64$ atoms using various alkali substrates. The potentials of interaction between helium atoms and the substrates, modeled as flat, are taken from Ref. \cite{Cole1998}. For this study, since the goal is that of gaining insight in the physics of an isolated cluster rather than bulk, the sides of the simulation cell are all taken large enough (of the order of several hundred \AA) to make periodic boundary conditions irrelevant.
It is worth noting that even for smallest clusters studied here, comprising $N=64$ atoms, there is no need for any artificial confining potential to keep the clusters together; for, clusters stay together at the temperature of these calculations, largely due to quantum-mechanical exchanges, known to impart stability to liquid-like superfluid phases \cite{Boninsegni2012b}. 
\\ \indent
Figs. \ref{dropCs} and \ref{dropLi} show radial density profiles integrated over the direction perpendicular to the substrate (i.e., effective 2D profiles) and computed with respect to an axis through the center of mass of the cluster, for two helium nanoclusters comprising $N=64$ atoms (one of $^3$He) adsorbed on Cs (top)  and Li (bottom) substrates. The first consideration to make is that despite the relatively small size, the shape of the cluster on the two substrates is markedly different. 
On the weaker (Cs) substrate $^4$He atoms ``bead up'' forming a nearly spherical droplet; the cluster is considerably more compact than that on a Li substrate, which instead consists of a single layer (albeit with significant atomic delocalization in the perpendicular direction), displays a nearly flat portion near the axis and extends out to a distance $\sim 60\%$ greater than on Cs. 
\\ \indent
These differences notwithstanding, in both cases {\em a}) the $^3$He atom binds to the cluster {\em b}) the bound state is sharply localized in the outer region of the cluster, near the substrate. This kind of bound state was first reported in  Ref. \cite{Mayol2003}, in which Density Functional calculations were carried out for droplets of $^4$He adsorbed on Cs. It is argued therein that only adsorbed droplets of sufficiently large size ($\sim 1,000$ $^4$He atoms) support such a bound state, suggesting that a nearly spherical $^4$He droplet with large enough a surface is required. 
\\ \indent
Our results show that a $^3$He bound state sharply localized at the edge exists even on much smaller (a fact already reported in Ref. \cite{Stipanovic2012} for a Cs substrate) and nearly 2D clusters, such as those adsorbed on Li; the failure to detect it in the original calculation of Ref. \cite{Mayol2003} may have to do with the fact that Density Functional approaches are largely built on bulk physics and may not capture the whole picture as the size of the system becomes small enough that surface physics may dominate over bulk.
\\ \indent
Fig. \ref{dropCs} shows a noticeable $^3$He penetration of the core region of the cluster, suggesting a possible, significant $^3$He delocalization with some interplay of surface and bulk physics. This might be the case on small clusters, but, as shown in Fig. \ref{dropCs2}, illustrating the result for a cluster with four times more atoms, the $^3$He bound state is confined to the edge, its width remarkably close to that observed for a 3D bulk surface ($\sim 10$\ \AA).
\begin{figure}[h]
\centering
     \includegraphics[width=1.0\linewidth]{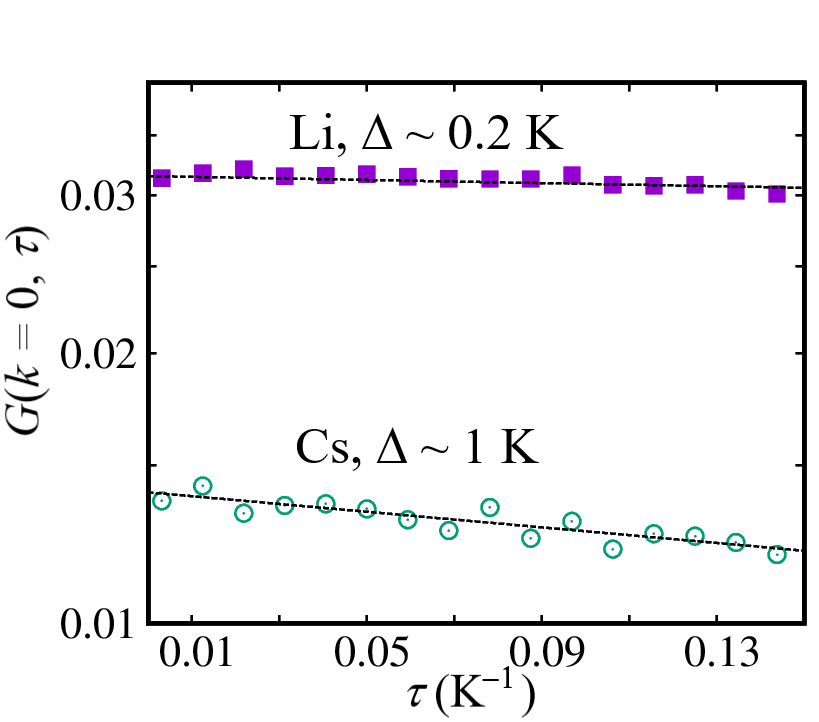}
\caption{Matsubara Green function $G({\bf k}=0,\tau)$ for a single $^3$He atom bound to a 63-atom $^4$He cluster adsorbed on Cs (open circles) and Li (filled squares) substrates (semi-logarithmic scale). Solid lines are exponential fits to the data, i.e., the fitting function is $\propto$ exp$[-\Delta\tau]$, where the fitting parameter $\Delta$ is the binding energy, measured with respect to the binding energy of a $^3$He atom to the substrate {\em in vacuo}. \label{goft}}
\end{figure} 
\\ \indent
The results on finite $^4$He droplets allow one to validate the proposition made above, based on simulations of 3D and 2D extended surfaces, that dimensional reduction acts to weaken the binding of $^3$He to a $^4$He surface. The binding energy of the $^3$He atom to a $^4$He cluster can be estimated from the decay in imaginary time of the low-temperature limit of the single-particle Matsubara Green function $G(k=0,\tau)$, which is an outcome of a QMC simulation based on the continuous-space Worm Algorithm (for details, see Ref. \cite{Boninsegni2006b}).
Fig. \ref{goft} shows this quantity for the $^3$He atom for the simulations whose results are shown in Figs. \ref{dropCs} and \ref{dropLi}. In both cases, the behavior is consistent with the exponential decay $\sim$ exp$[-\Delta\tau]$, where $\Delta$ is the binding energy measured with respect to the binding energy of a single $^3$He atom to the substrate, i.e., with no $^4$He cluster present (this quantity is worth 3.46 K for Cs, 9.72 K for Li with the atom-substrate interactions utilized in this work). The faster decay (i.e., stronger binding) for a Cs substrate, on which the droplet is more markedly 3D, is clear; although the estimation of the binding energy is not very precise, nonetheless a factor $\gtrsim 5$ difference in the relative strength of the binding on the two substrates can easily be established.
In order to make this conclusion more definitive, simulations were also carried out for a 64-atom cluster adsorbed on a fictitious flat substrate with the strength of graphite, described by the laterally averaged Carlos-Cole potential \cite{Carlos1980}. In this case, the strength of the substrate causes the $^4$He cluster to be very nearly 2D, and no evidence of binding of a $^3$He atom is observed, at least down to the temperature considered here, namely $T=0.125$ K. All of this is consistent with the discussion of the results of the previous subsections.
\\ \indent
Before concluding this subsection, it is worth noting that, while its binding energy is considerably reduced with respect to that on Cs, a bound state exists on a Li substrate as well, for a $^4$He cluster of this size, and therefore {\em a fortiori} 
on all remaining alkali substrates, which are less attractive than Li.
\\
\section{Conclusions}\label{concl}
Results of extensive QMC simulations of a single $^3$He atom at a free $^4$He surface in different physical settings confirm the existence of a bound state at a 3D superfluid $^4$He surface. The bound state is localized within a $\sim$ 10 \AA\ thick low-density $^4$He surface layer,  with a binding energy of approximately 5 K  with respect to vacuum. This estimate is consistent with the most recent available one, obtained by Density Functional calculations \cite{Bashkin1995}. A possible surface bound state of two $^3$He atoms (i.e., a ($^3$He)$_2$ dimer)  is of great interest, given its relevance to the fascinating scenario of a quasi-2D $^3$He superfluid phase floating on top of superfluid $^4$He. Such a bound state can in principle be investigated using the same methodology adopted in this work; preliminary simulation results obtained at temperature  $T=250$ mK suggest that the dimer, if it exists, is weakly bound, i.e, its size likely exceeds significantly that (14.3 \AA) of our simulation cell in the direction perpendicular to the surface. Thus, its unambiguous, quantitative characterization requires simulations on considerably larger system sizes (chiefly larger surfaces) than that on which the results reported here were obtained. It remains therefore an ongoing project at this time. 
\\ \indent
It is found that if the dimensionality of the system is lowered no $^3$He bound state exists, either dissolved in 2D bulk superfluid $^4$He nor at a linear $^4$He surface. This fact, which appears not to have been investigated before, clearly raises doubts about the relevance of previous studies of helium mixtures in two dimensions; it shows that the well-known property of superfluid $^4$He of  ``cleaning'' itself of impurities is strengthened in reduced dimensions, extending to its lighter $^3$He isotope, i.e., the only atom that can be dissolved in superfluid $^4$He in three dimensions. This observation can in principle be verified experimentally with small helium clusters adsorbed on weak substrates. Simulations carried out in this work for such systems confirm the existence of bound states of $^3$He atoms at the edge of nanoscale $^4$He droplets adsorbed on {\em all} alkali metal substrates, i.e., not only Cs; binding is weakened as the attractive strength of the substrate is increased, and consequently the $^4$He droplet takes on an increasingly 2D-like shape.
\\ \indent
This work was supported by the Natural Sciences and Engineering Research Council of Canada, under grant RGPIN 2024-05664. 
\bibliography{refs}

\end{document}